\documentclass[aps,prl,twocolumn,superscriptaddress]{revtex4}
\usepackage{graphicx}

\begin{document} 

\title{A Measurement of the $K^0$ Charge Radius and a CP Violating Asymmetry
Together with a Search for CP Violating E1 Direct Photon Emission in 
the Rare Decay $K_{L}\rightarrow\pi^{+}\pi^{-}e^{+}e^{-}$}

%\draft

\author{
E.~Abouzaid$^{4}$,
%T.~Alexopoulos$^{12}$,
M.~Arenton$^{11}$,
%R.F.~Barbosa$^{10}$,
A.R.~Barker$^5$,
L.~Bellantoni$^7$,
A.~Bellavance$^{9}$,
E.~Blucher$^4$,
G.J.~Bock$^7$,
E.~Cheu$^1$,
R.~Coleman$^7$,
M.D.~Corcoran$^{9}$,
G.~Corti$^{11}$,
B.~Cox$^{11}$,
A.R.~Erwin$^{12}$,
C.O.~Escobar$^{3}$,
A.~Glazov$^4$,
A.~Golossanov$^{11,\dagger}$,
R.A.~Gomes$^{4}$,
P.~Gouffon$^{10}$,
K.~Hanagaki$^8$,
Y.B.~Hsiung$^7$,
H.~Huang$^5$,
D.A.~Jensen$^7$,
R.~Kessler$^4$,
K.~Kotera$^8$,
A.~Ledovskoy$^{11}$,
P.L.~McBride$^7$,
E.~Monnier$^{4,*}$,
K.S.~Nelson$^{11}$,
H.~Nguyen$^7$,
R.~Niclasen$^5$,
H.~Ping$^{12}$,
X.R.~Qi$^7$,
E.J.~Ramberg$^7$,
R.E.~Ray$^7$,
M. Ronquest$^{11}$,
E.~Santos$^{10}$,
J.~Shields$^{11}$,
W.~Slater$^2$,
D.~Smith$^{11}$,
N.~Solomey$^4$,
E.C.~Swallow$^{4,6}$,
P.A.~Toale$^5$,
R.~Tschirhart$^7$,
C.~Velissaris$^{12}$,
Y.W.~Wah$^4$,
J.~Wang$^1$,
H.B.~White$^7$,
J.~Whitmore$^7$,
M.~Wilking$^5$,
B.~Winstein$^4$,
R.~Winston$^4$,
E.T.~Worchester$^4$,
M.~Worchester$^4$,
T.~Yamanaka$^8$,
E.D.~Zimmerman$^5$,
R.F.~Zukanovich$^{10}$\\
(KTeV Collaboration)} 

%\address{\vspace{0.1in}
\affiliation{
%\noindent 
$^{1}$University of Arizona, Tucson, Arizona 85721 \\
$^{2}$University of California at Los Angeles, Los Angeles, California 90095\\
$^{3}$Universidade Estadual de Campinas, Campinas, Brazil 13083-970\\
$^{4}$The Enrico Fermi Institute, The University of Chicago, Chicago, Illinois
60637\\
$^{5}$University of Colorado, Boulder Colorado 80309\\
$^{6}$Elmhurst College, Elmhurst, Illinois 60126\\
$^{7}$Fermi National Accelerator Laboratory, Batavia, Illinois 60510\\
$^{8}$Osaka University, Toyonaka, Osaka 560 Japan\\
$^{9}$Rice University, Houston, Texas 77005\\
$^{10}$Universidade de Sao Paulo, Sao Paulo, Brazil 05315-970\\
%$^{11}$The Dept. of Physics and Institute of Nuclear and Particle
%Physics, University of Virginia, Charlottesville, Virginia 22901\\
$^{11}$University of Virginia, Charlottesville, Virginia 22901\\
$^{12}$University of Wisconsin, Madison, Wisconsin 53706\\ }

\begin{abstract} 
%\myabstract{
Using the complete KTeV data set of 5241 candidate
$K_{L}\rightarrow\pi^{+}\pi^{-}e^{+}e^{-}$ decays (including an estimated background of
204$\pm$14 events), we have measured the coupling $g_{CR}=0.163\pm 0.014~({\rm stat})\pm
0.023~({\rm syst})$ of the CP conserving charge radius process
and from it determined a $K^0$ charge radius of $\langle r^2_{K^0} \rangle =
(-0.077\pm0.007({\rm stat})\pm0.011({\rm syst})) fm^2$. We have also determined a first experimental upper
limit of 0.04 (90\% CL) for the ratio $\frac{|g_{E1}|}{|g_{M1}|}$
of the coupling for the E1 direct photon emission process
relative to the coupling for M1 direct photon emission process. We also report the
measurement of $|g_{M1}|$ including its associated vector form factor 
$|\tilde{g}_{M1}|(1+ \frac{a_1/a_2}{(M^2_{\rho}-M^2_K)+2M_KE_{\gamma^*}})$
where $|\tilde{g}_{M1}| = 1.11\pm{0.12}~({\rm stat})\pm 0.08~({\rm syst})$
and $a_1/a_2 = (-0.744\pm0.027~({\rm stat})\pm 0.032~({\rm syst})) GeV^2/c^2$. In addition, a measurement of the 
manifestly CP violating asymmetry of magnitude $(13.6\pm 1.4~({\rm stat})\pm 1.5~({\rm syst}))$\% 
in the CP and T odd angle $\phi$ between the decay planes of the $e^{+}e^{-}$ and $\pi^{+}\pi^{-}$
pairs in the $K_L$ center of mass system is reported.\\ \\
\vspace{0.25in}
%}
\end{abstract}

% body of paper here
\maketitle

The emission of a virtual photon in the rare decay $K_{L}\rightarrow\pi^{+}\pi^{-}e^{+}e^{-}$ proceeds via 
three main processes: bremsstrahlung, direct photon emission, and the charge
radius process. The bremsstrahlung process takes place via the CP violating decay of a
$K_L\rightarrow\pi^{+}\pi^{-}$ followed by emission of an electric dipole
(E1) photon by bremsstrahlung from one of the $\pi$'s. The direct emission
process involves either the CP conserving or CP violating direct emission at 
the primary decay vertex of a magnetic dipole (M1) or a electric dipole
(E1) photon respectively. The CP conserving charge radius process is the 
transformation of a $K_L\rightarrow K_S$ by emission of a virtual
photon in a J=0 transition (forbidden in real photon emission) 
followed by the CP conserving decay of 
the $K_S$ into $\pi^+\pi^-$. The charge radius coupling is related to the charge
radius of the neutral kaon since the virtual photon acts as a probe of the
$K^0$ in a way similar to the virtual photon in $K^0$ scattering from an atomic electron.
The E1 and M1 direct emission and charge radius couplings
are $g_{E1}$, $g_{M1}$ and $g_{CR}$. The matrix elements~\cite{ref:4} for 
the bremsstrahlung, M1, E1 and charge radius processes are 

\begin{eqnarray}
\raggedleft
M_{br}&\sim & \eta_{+-}e^{i\delta_0(M^2_K)}\left[\frac{p_{+\mu}}{p_{+}\cdot k}-
\frac{p_{-\mu}}{p_{-}\cdot k}\right]\frac{\overline{u}(k_-)\gamma^{\mu}v(k_+)}{k^2} \nonumber \\ 
M_{M1}&\sim &i|g_{M1}|e^{i\delta_1(M^2_{\pi\pi})}\epsilon_{\mu\nu\rho\sigma}k^{\nu}p^{\rho}_+p^{\sigma}_-
\frac{\overline{u}(k_-)\gamma^{\mu}v(k_+)}{k^2} \\
M_{E1}&\sim &|g_{E1}|e^{i(\phi_{+-}+\delta_1(M^2_{\pi\pi}))}[(p_-\cdot
k)p_{+\mu}-(p_+\cdot k)p_{-\mu}] \nonumber \\
& &\cdot \frac{\overline{u}(k_-)\gamma^{\mu}v(k_+)}{k^2} \nonumber \\
M_{CR}&\sim &|g_{CR}|e^{i\delta_0(M^2_{\pi\pi})}\frac{k^2 P_{\mu}-(P\cdot
k)k_{\mu}}{M^2_{\pi\pi}-M^2_K}
\frac{\overline{u}(k_-)\gamma^{\mu}v(k_+)}{k^2} \nonumber
\end{eqnarray}

\noindent where $p_+$, $p_-$, $k_+$, $k_-$, k, P are the $\pi^+$, $\pi^-$,
positron, electron, virtual photon, and $K_L$ four
momenta. The $\delta_{0,1}$ are the I=0,1 $\pi^+\pi^-$ strong interaction
phase shifts. $\eta_{+-}$ is the coupling of the $K_L\rightarrow\pi^+\pi^-$ decay.

The KTeV E799-II experiment at Fermi National Accelerator Laboratory
previously reported the first observation~\cite{ref:1} of the rare four body decay 
mode $K_{L}\rightarrow\pi^{+}\pi^{-}e^{+}e^{-}$ based on 1\% of the KTeV
data.  We have also made an initial measurement~\cite{ref:2} based on 36\% of the KTeV 
$K_{L}\rightarrow\pi^{+}\pi^{-}e^{+}e^{-}$ data of a CP-violating asymmetry 
in the variable $\sin\phi\cos\phi$ (where $\phi$ is the CP- and T-odd angle 
between the $e^+e^-$ and $\pi^+\pi^-$ planes in the $K_L$ cms). In addition, 
the measurement of the M1 direct photon emission coupling $|g_{M1}|$ 
including a vector form factor was reported in Ref.~\cite{ref:2}. 
In this paper we report a measurement of the charge radius of the $K^0$ 
obtained from the coupling $|g_{CR}|$ of the charge radius process of the
$K_{L}\rightarrow\pi^{+}\pi^{-}e^{+}e^{-}$ decay. We also determined an 
upper limit for the E1 direct photon emission in this decay.
Finally, we present the measurements of the M1 direct emission process 
coupling and its form factor and the CP violating asymmetry in $\sin\phi\cos\phi$.

\begin{figure}[h!]
 \begin{center}
\includegraphics[width=0.295\textwidth]{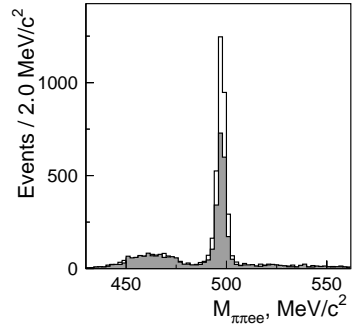}
%\epsfxsize=0.70\hsize
%\epsfbox{ppee_mass_asym_gray.eps}
 \end{center}
\caption{$M_{\pi^{+}\pi^{-}e^{+}e^{-}}$ invariant mass 
for events passing all $K_{L}\rightarrow\pi^{+}\pi^{-}e^{+}e^{-}$ physics cuts. 
The superimposed $K_L$ mass peaks 
for $\sin\phi\cos\phi >0$ (white histogram)  and $<0$ (gray histogram) directly demonstrate the
large CP violating asymmetry.  There is no asymmetry in the backgrounds to
the two peaks as demonstrated by their complete overlap of the distributions
outside the kaon peak region.} 
\label{Fig:3}
\end{figure}

The $K_{L}\rightarrow\pi^{+}\pi^{-}e^{+}e^{-}$ data were accumulated 
during the 1997 and 1999 runs of the KTeV E799-II experiment. Differences
in running conditions and spectrometer configuration can be found in
Ref.~\cite{ref:13}. The total KTeV E799-II $K_{L}\rightarrow\pi^{+}\pi^{-}e^{+}e^{-}$ 
signal of $5241$ events, including an estimated background of
$204\pm 14$ events, obtained after the analysis cuts described below, 
is shown in the $\pi^+\pi^-e^+e^-$ mass plot of Fig.~\ref{Fig:3}.
Note that data has been separately plotted for $\sin\phi\cos\phi~>0$ and
$\sin\phi\cos\phi~<0$.  The CP violating asymmetry can be seen directly in 
the mass plot in the differing sizes of the two mass peaks.

The KTeV four track trigger~\cite{ref:2} selected 
$3.9\times 10^8$ events from the 97 and 99 runs. Candidate 
$K_{L}\rightarrow\pi^{+}\pi^{-}e^{+}e^{-}$ events 
were extracted from these triggers by requiring 
events to have four tracks that passed track quality cuts and had 
a common vertex with a good vertex $\chi^2$.  
To be designated as $e^{\pm}$, two of the tracks were required 
to have opposite charges and $0.95\leq {\rm E/p}\leq~1.05$, 
where E was the energy deposited by the track in the calorimeter,
and p was the momentum obtained from magnetic deflection.  
To be consistent with a $\pi^{\pm}$ pair, the other two tracks were 
required to have ${\rm E/p}\leq 0.90$ and opposite charges.  
To reduce backgrounds arising from other types of $K_L$ decays in which decay products 
have been missed, the candidate $\pi^{+}\pi^{-}e^{+}e^{-}$ 
were required to have transverse momentum $P_{t}^{2}$ of the four tracks
relative to the direction of the $K_{L}$ be less than 
$0.6\times 10^{-4}~{\rm GeV}^{2}/c^{2}$.  This cut was 94\% efficient
for retaining $K_{L}\rightarrow\pi^{+}\pi^{-}e^{+}e^{-}$.

The major background to the $K_{L}\rightarrow\pi^{+}\pi^{-}e^{+}e^{-}$ mode was
$K_{L}\rightarrow\pi^{+}\pi^{-}\pi^{0}_D$ where $\pi^{0}_D$ was a 
Dalitz decay, $\pi^{0}\rightarrow\gamma e^+e^-$, in which the photon
was not observed in the CsI calorimeter or the photon vetos.  To reduce this background, all 
$K_{L}\rightarrow\pi^{+}\pi^{-}e^{+}e^{-}$ candidate events were
interpreted as $K_{L}\rightarrow\pi^{+}\pi^{-}\pi^{0}_D$ decays.
Under this assumption, the longitudinal momentum squared $(P^2_L)_{\pi^0}$ of the
assumed $\pi^0$ can be calculated in the frame in which the 
momentum of $\pi^{+}\pi^{-}$ is transverse to the $K_L$ direction.
$(P^2_L)_{\pi^0}$ was greater than zero for
$K_{L}\rightarrow\pi^{+}\pi^{-}\pi^{0}_D$ decays except for cases where
finite detector resolution resulted in a $(P^2_L)_{\pi^0}\leq~0$.
In contrast, most of the $K_{L}\rightarrow\pi^{+}\pi^{-}e^{+}e^{-}$ decays 
had $(P^2_L)_{\pi^0}\leq~0$.  The requirement that all 
$\pi^{+}\pi^{-}e^{+}e^{-}$ had 
$(P^2_L)_{\pi^0} \leq -0.025\ {\rm GeV}^{2}/c^{2}$
reduced the $K_{L}\rightarrow\pi^{+}\pi^{-}\pi^{0}_D$ background under the
$K_L$ peak to 177 events while retaining 94\% of the signal. 

A second significant background was due to $\Xi^0\rightarrow\Lambda\pi^0\rightarrow
p\pi^-e^+e^-\gamma$ where the photon was missed and the proton was misidentified
as a $\pi^+$.  There were 22 events of background after all cuts due to this decay.
All other backgrounds were relatively minor. The largest was due
to $K_L\rightarrow\pi^0\pi^0\pi^0$ with $\pi^0\rightarrow e^+e^-e^+e^-$.
This mode contributed approximately four events to the background after cuts.  In
addition, a potentially large background due to $K_{L}\rightarrow\pi^{+}\pi^{-}\gamma$ 
decays in which the photon converted in the material of the spectrometer
producing an $e^+e^-$ pair was eliminated by requiring 
$M_{e^{+}e^{-}}\geq 2.0\ {\rm MeV}/c^{2}$.  The $M_{e^{+}e^{-}}$ cut 
retained 95\% of the $K_{L}\rightarrow\pi^{+}\pi^{-}e^{+}e^{-}$ events 
with only one event contributing to the background.

The final requirement of the $K_{L}\rightarrow\pi^{+}\pi^{-}e^{+}e^{-}$ events was 
that $492~{\rm MeV}/c^2\leq M_{\pi\pi ee}\leq 504~{\rm MeV}/c^2$.  The magnitude of the 
background under the $K_{L}$ peak was determined by a fit to the
simulated background mass distribution to the wings of the signal region. 
From this fit, a $K_{L}\rightarrow\pi^{+}\pi^{-}e^{+}e^{-}$ signal of
$5037$ events above a background of $204\pm 14$ events was obtained. 

\begin{figure}[h!] 
 \centering
 \includegraphics[width=3.5cm]{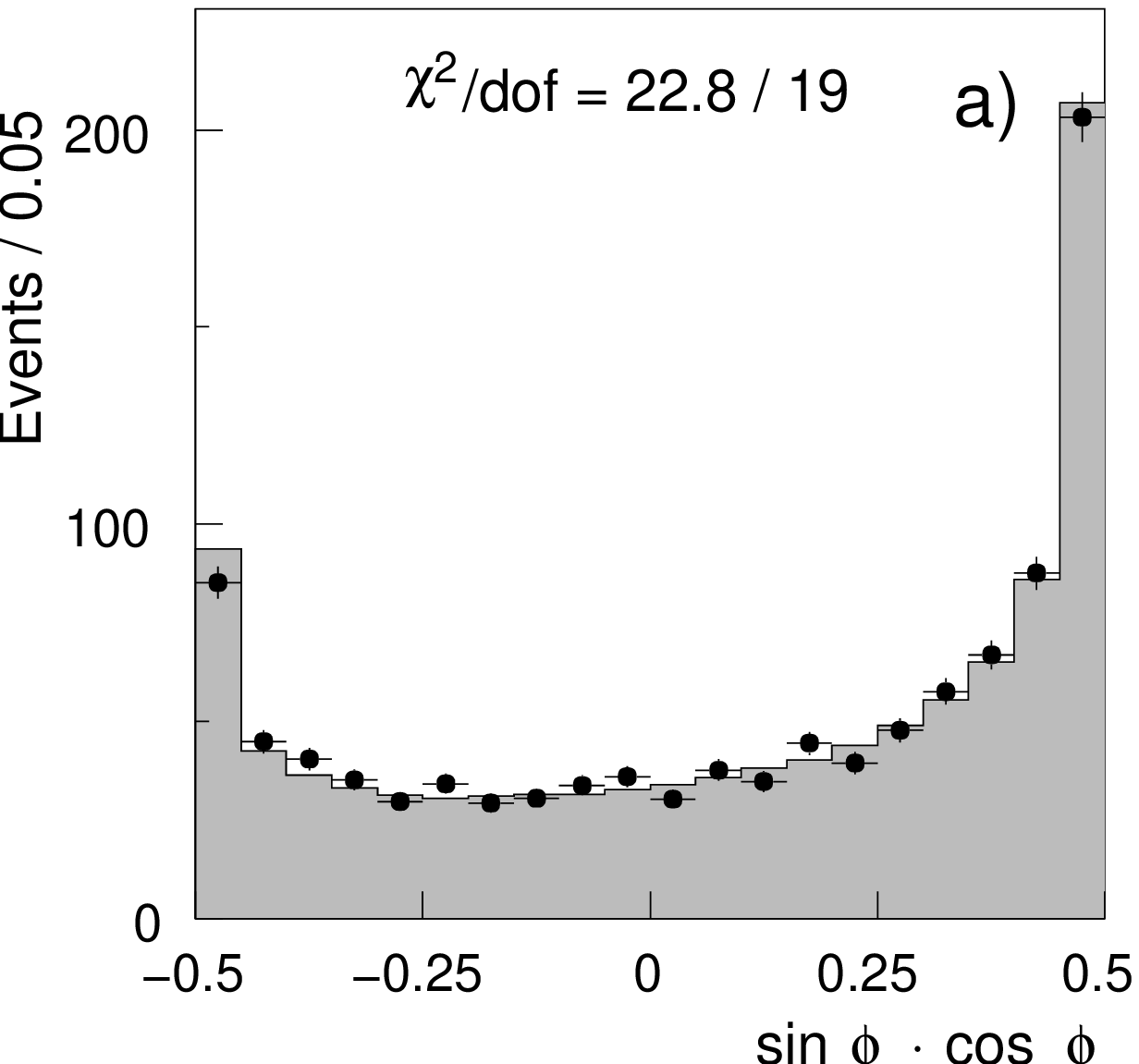}
 \includegraphics[width=3.5cm]{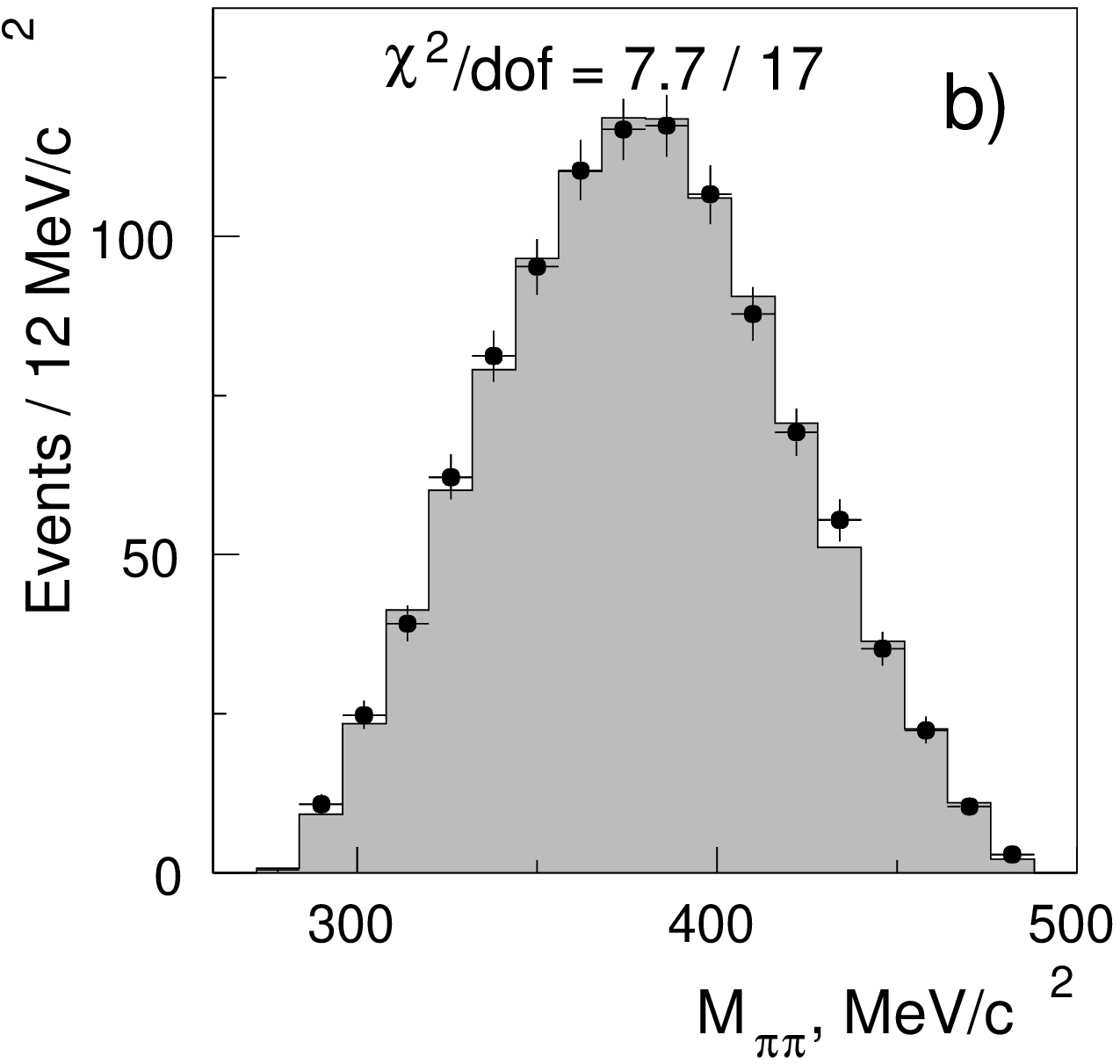}\\
 \includegraphics[width=3.5cm]{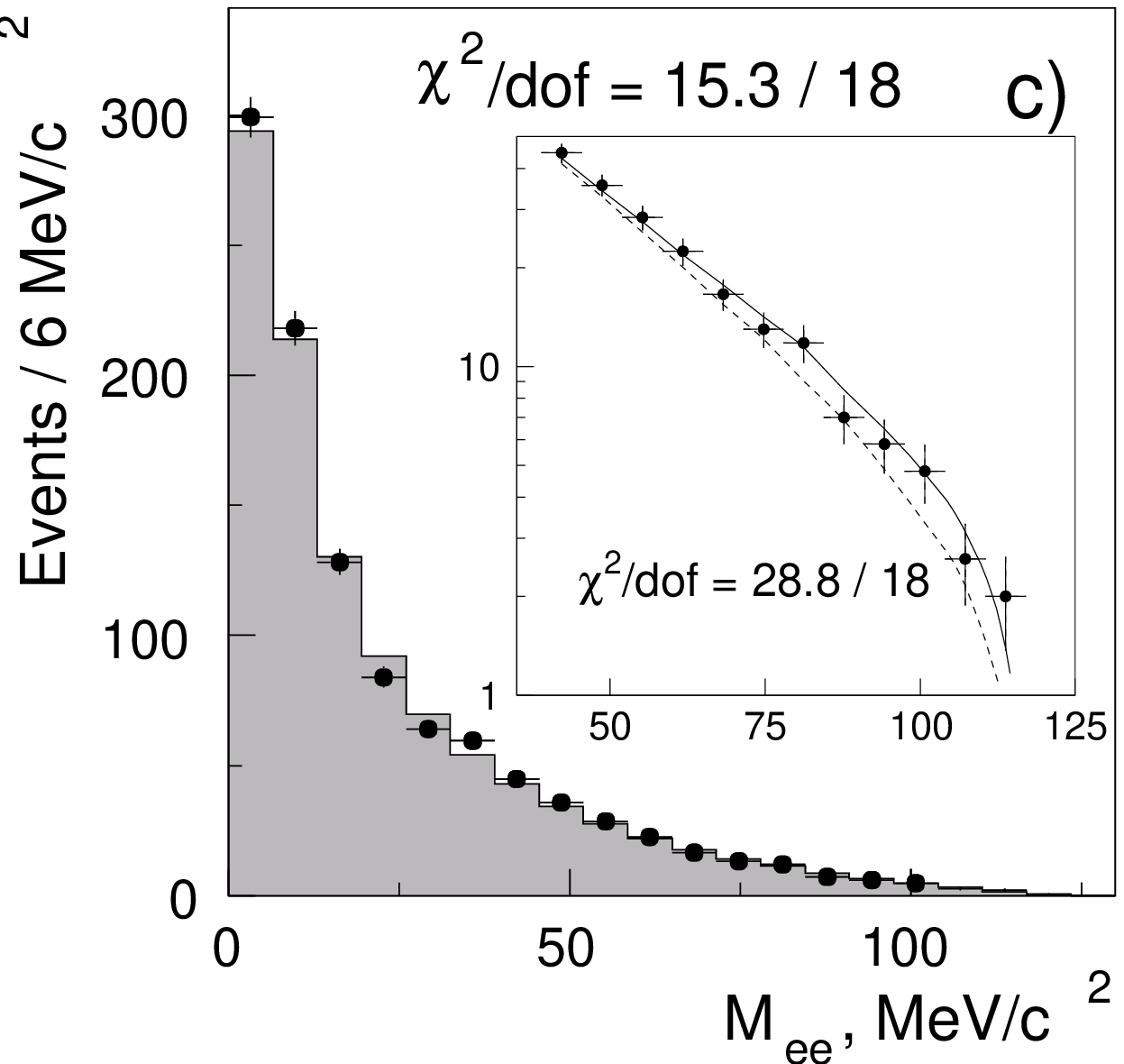}
 \includegraphics[width=3.5cm]{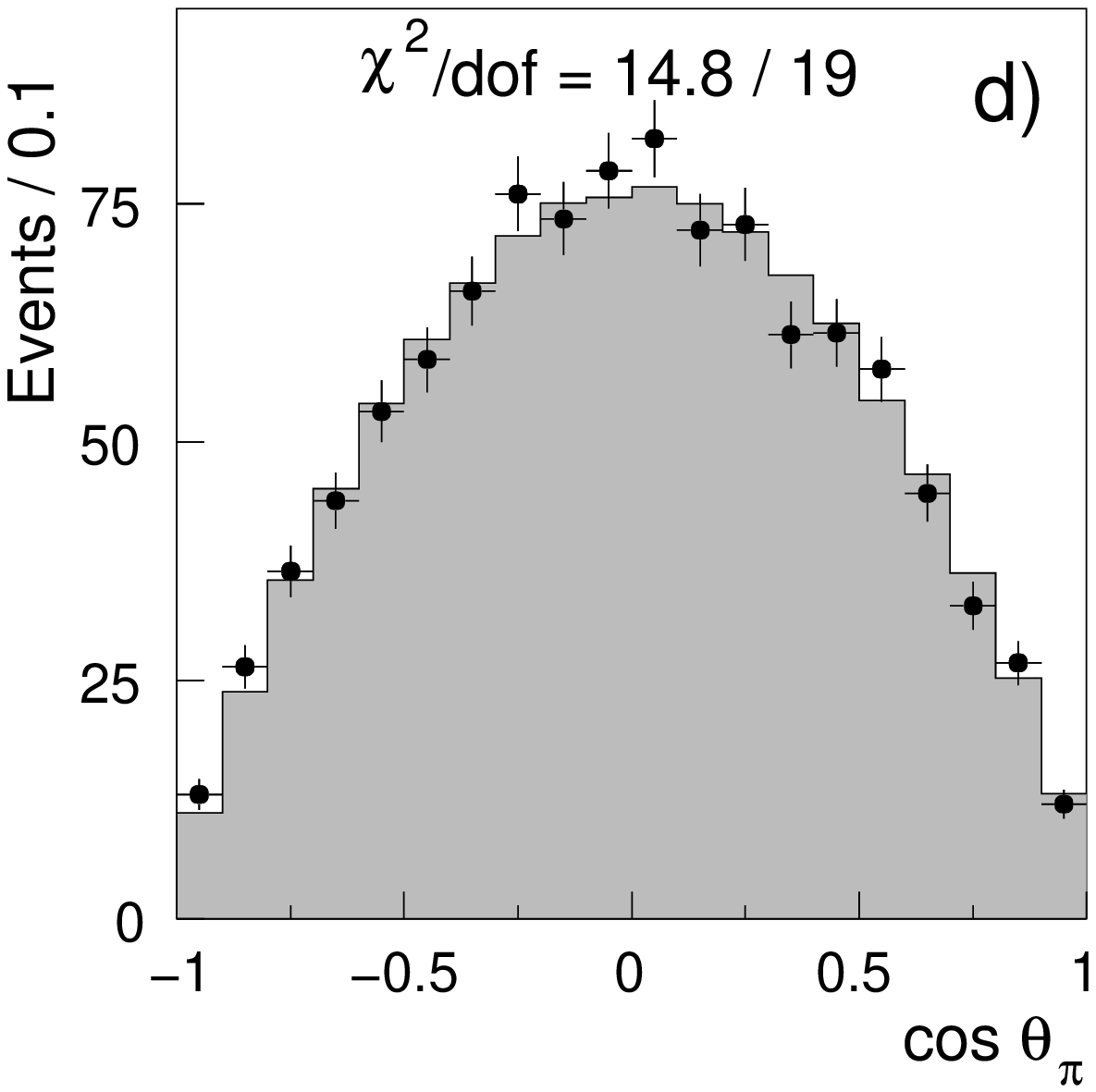}\\
 \includegraphics[width=3.5cm]{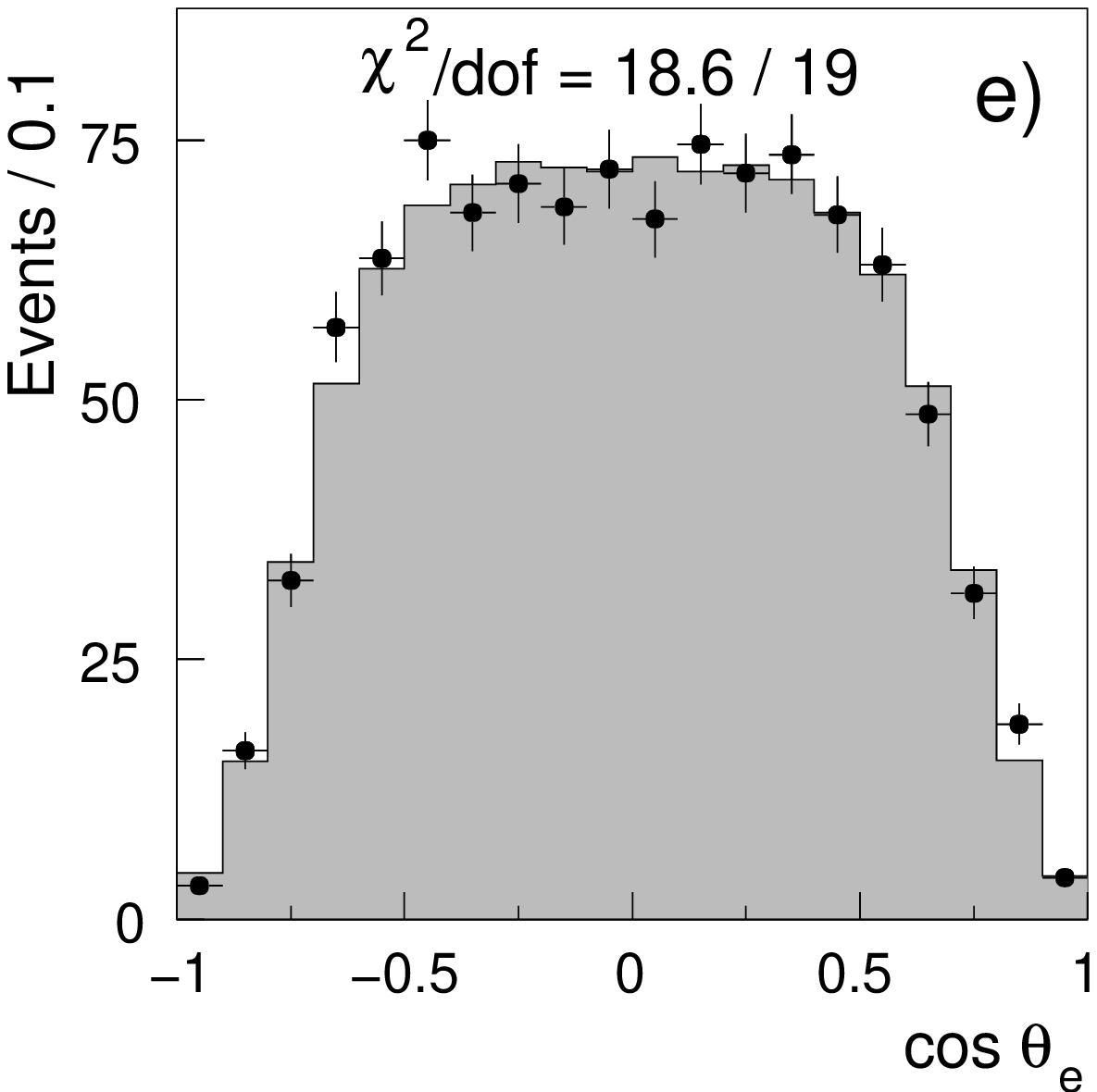}
 \includegraphics[width=3.5cm]{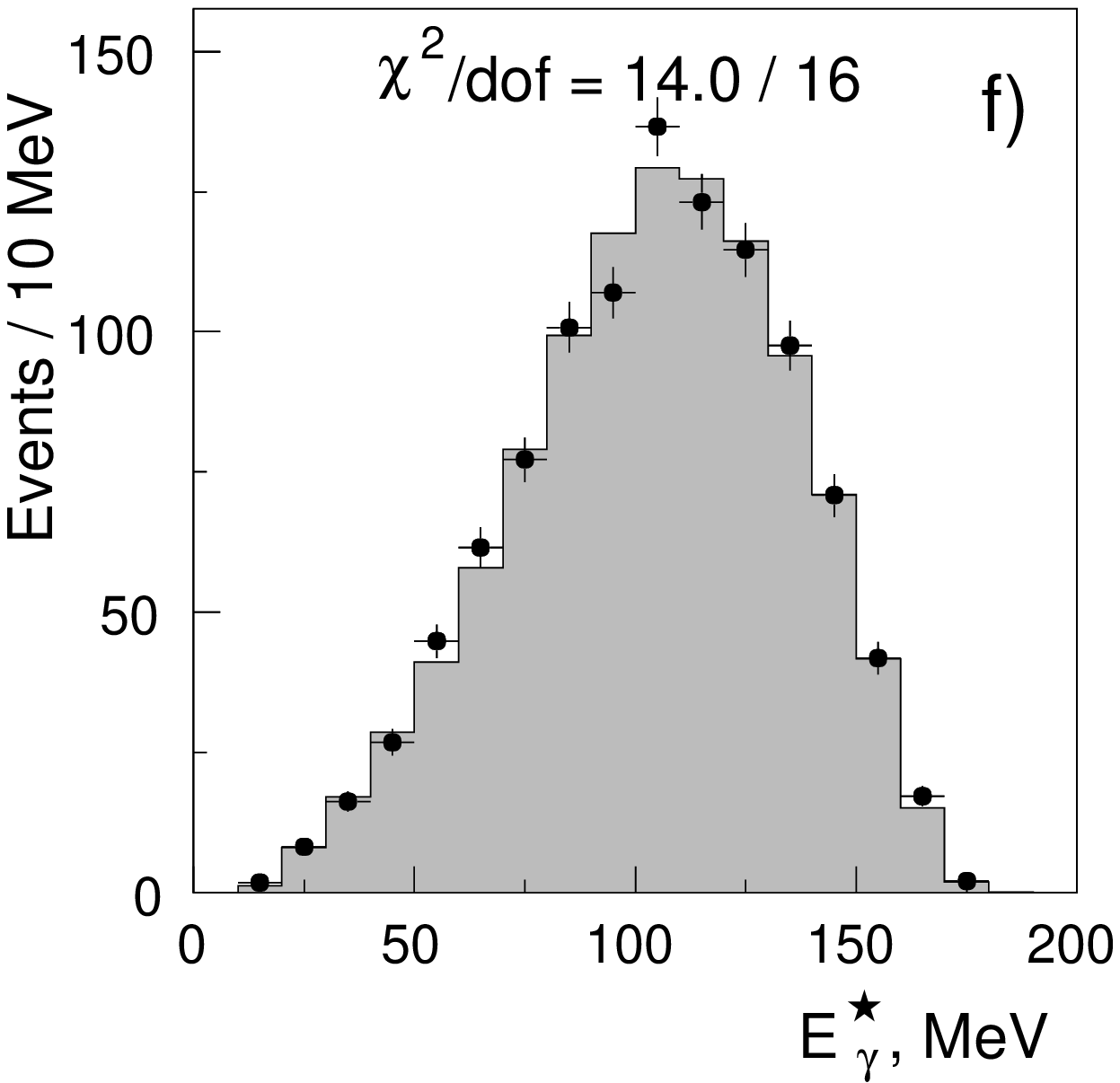}
\caption{Likelihood fit to the five independent variables
a) $sin\phi$$cos\phi$, b) $M_{\pi^+\pi^-}$, c)  $M_{e^{+}e^{-}}$ 
(the dotted curve in the insert in this figure shows the 
deficit of $e^+e^-$ pairs at high $M_{ee}$ if the charge radius process is 
ignored.  The $\chi^2/dof$ for the fit of the model to the data increases from
0.85 to 1.6 if the charge radius process is left out.),
d) $\theta_{\pi^{+}}$, and e) $\theta_{e^{+}}$ of the $K_{L}\rightarrow\pi^{+}\pi^{-}e^{+}e^{-}$
decay; f)  $E_{\gamma^*}$ defined as $E_{e^+}+E_{e^-}$ }
\label{Fig:10}
\end{figure}

We analyzed the $K_{L}\rightarrow\pi^{+}\pi^{-}e^{+}e^{-}$ decays
in a likelihood fit that used the matrix elements (eq. 1) of the 
model of~\cite{ref:4} with additional radiative corrections 
applied to the final state particles using the PHOTOS program~\cite{ref:6a}. 
We also found it necessary to include a dependence on the virtual photon energy
in the M1 virtual photon emission coupling in order to obtain 
agreement with the virtual photon energy spectrum $E_{\gamma^*}=E_{e^+}+E_{e^-}$ 
of the data (Fig.~\ref{Fig:10}f).  The M1 coupling $|g_{M1}|$
was modified by a form factor

\begin{equation}
|g_{M1}|=|\tilde{g}_{M1}| \left[1+ \frac{a_1/a_2}{(M^{2}_{\rho}-M^{2}_{K})+2M_{K}E_{\gamma^*}}\right]
\end{equation}

\noindent similar to that used in Ref~\cite{ref:6} to describe $K_{L}\rightarrow\pi^{+}\pi^{-}\gamma$.
Here $M_{\rho}$ is the mass of the $\rho$ meson (770 {\rm MeV}/$c^2$) and the 
photon energy has been replaced by $E_{e^{+}}+E_{e^{-}}$. 

%The dominant processes were the CP violating bremsstrahlung and 
%CP-conserving M1 direct photon emission. The CP-conserving 
%charge radius process also contributed significantly while the
%CP-violating E1 direct photon emission was too small to detect.

%\begin{figure}[h!]
% \begin{center}
%\epsfxsize=0.80\hsize
%\epsfbox{amplitudes_10.eps}
% \end{center}
%\caption{Processes contributing to
%$K_{L}\rightarrow\pi^{+}\pi^{-}e^{+}e^{-}$. a) 
%CP-violating bremsstrahlung b) CP-conserving M1 $\gamma$ emission
%c) CP-violating E1 $\gamma$ emission d) Charge radius process}
%\label{Fig:1}
%\end{figure}

The likelihood of a given event (see eq. 3 below), based on the matrix
elements $\mu(\vec{x}_i,\vec{\alpha})$ of the model of
Ref.~\cite{ref:4}, is a function of the five independent variables $\vec{x}_i$: 
$\phi$, $\theta_{e^{+}}$ (the angle between the $e^{+}$ and the $\pi^{+}\pi^{-}$ direction in 
the $e^{+}e^{-}$ cms), $\theta_{\pi^{+}}$ (the angle between the $\pi^{+}$ 
and the $e^{+}e^{-}$ direction in the $\pi^{+}\pi^{-}$ cms), $M_{\pi^{+}\pi^{-}}$, 
and $M_{e^{+}e^{-}}$.  In addition, it depends on the values of the fit
parameters $\vec{\alpha}$: $a_1$/$a_2$ and $|\tilde{g}_{M1}|$, $\frac{|g_{E1}|}{|g_{M1}|}$, $|g_{CP}|$ and 
nominal values for other model parameters such as
$\eta_{+-}=(2.286\pm0.017)\times10^{-3}$ 
and $\Phi_{+-}=43.51^0\pm0.06^0$.  
The measured strong interaction phase shifts of the $\pi^+\pi^-$ were
taken from Ref.~\cite{ref:phases}.  The likelihood was calculated using the
selected $K_L\rightarrow\pi^+\pi^-e^+e^-$ data sample of $N_D$ events and a large 
Monte Carlo event sample $N_{MC}$ generated with nominal values of the
fit parameters $\vec{\alpha}_0$, passed through the spectrometer and reconstructed, and then reweighted
with a new set of fit parameters $\vec{\alpha}$ using the matrix elements $\mu(\vec{x}_i,\vec{\alpha})$. 
The likelihood fit to the five independent variables is shown in 
Fig.~\ref{Fig:10} along with the fit to $E_{\gamma^*}$. The charge radius process contributes to the
higher mass $M_{ee}$ (as shown in the insert of Fig.~2c) while the M1
direct emission is determined by the shape of the $M_{\pi^+\pi^-}$ spectrum.

The likelihood function used to perform the fit is

\begin{equation}
lnL(\vec{\alpha})=\sum_{i=1}^{N_D} ln\mu (\vec{x}_i,\vec{\alpha})-N_{D}ln\sum_{j=1}^{N_{MC}} \frac{\mu
(\vec{x}_j,\vec{\alpha})}{\mu (\vec{x}_j,\vec{\alpha}_0)}
\end{equation}

\noindent The best fit values were $a_1$/$a_2 = (-0.744\pm0.027({\rm
stat}))~{\rm GeV}^2/c^2$, $|\tilde{g}_{M1}| = 1.11\pm0.12(stat)$, 
$|g_{CP}|=0.163\pm0.014({\rm stat})$ and  $\frac{|g_{E1}|}{|g_{M1}|}$ $\leq 0.028$ (upper limit
due to statistical uncertainty only). The correlation ($\rho=0.924$) between $a_1$/$a_2$
and $|\tilde{g}_{M1}|$ has been taken into account in determining their errors.

The distribution of the quantity $\sin\phi\cos\phi$ 
(given by $(\hat{n}_{ee}\times \hat{n}_{\pi\pi})\cdot\hat{z}(\hat{n}_{ee}\cdot\hat{n}_{\pi\pi})$,
where the $\hat{n}'s$ are the unit normals to the ee and $\pi\pi$ planes and $\hat{z}$ is the
unit vector in the $\pi\pi$ direction in the $K_L$ cms) is shown in
Fig.~\ref{Fig:10}a.  The asymmetry of the $\sin\phi\cos\phi$ distribution

\begin{equation}
A = \frac{N_{\sin\phi \cos\phi~>~0.0} -
N_{\sin\phi \cos\phi~<~0.0}}
{N_{\sin\phi \cos\phi~>~0.0}+N_{\sin\phi \cos\phi~<~0.0}}
\end{equation}

\noindent yields $(23.8\pm{1.4}~({\rm stat}))$\% before acceptance corrections.  Using 
the fit of the model of Ref.~\cite{ref:4} to the data to determine the
acceptance, an asymmetry integrated 
over the entire $K_{L}\rightarrow\pi^{+}\pi^{-}e^{+}e^{-}$ phase space
of $(13.6\pm 1.4~({\rm stat}))$\% was obtained, the largest such CP violating
effect yet observed in kaon decay.  
The interference between the M1 direct emission process and the  
bremsstrahlung process generates the asymmetry in the  $\sin\phi\cos\phi$ distribution.  

Possible sources of false asymmetries were considered including
those due to backgrounds and asymmetries in the detector.
To check for detector asymmetries, a sample of $15\times 10^6$
$K_{L}\rightarrow\pi^{+}\pi^{-}\pi^{0}_D$ decays, which are expected to 
have no $\phi$ asymmetry and which have similar topology to $K_{L}\rightarrow\pi^{+}\pi^{-}e^{+}e^{-}$
except for the presence of an extra photon in the CsI, was examined, and an 
asymmetry of $((3.3\pm2.6)\times10^{-2}$)\% was measured. 

Systematic errors on $a_1$/$a_2$, $|\tilde{g}_{M1}|$, $|g_{CR}|$ and  $\frac{|g_{E1}|}{|g_{M1}|}$
due to several sources are shown in Table~\ref{systematics} below. As shown 
 
%\begin{figure}[h!]
% \begin{center}
%\epsfxsize=0.9\hsize
%\epsfbox{six_4.1.eps}
% \end{center}
%\caption{Likelihood fit to the five independent variables
%a) $\phi$, b) $\theta_{e^{+}}$, c) $\theta_{\pi^{+}}$, d) $M_{\pi^+\pi^-}$, and e) $M_{e^{+}e^{-}}$  
%of the $K_{L}\rightarrow\pi^{+}\pi^{-}e^{+}e^{-}$ decay;  f)Comparison of MC to $E^*_{\gamma}$ }
%\label{Fig:10}
%\end{figure}

%\begin{figure}[h!]
% \begin{center}
%\epsfxsize=0.50\hsize
%\epsfbox{dom_sincos_gray_2.eps}
% \end{center}
%\caption{a) Observed  asymmetry in the $\sin\phi \cos\phi$ angular distribution:
%The data are shown as dots. The histogram is a Monte Carlo simulation 
%based on the model of Refs. [3,4].}
%\label{Fig:5}
%\end{figure}

\noindent in Table~\ref{systematics}, the dominant systematic error is 
due to the variation of the fitted parameters
resulting from varying the physics cuts used to select the 
$K_{L}\rightarrow\pi^{+}\pi^{-}e^{+}e^{-}$data and Monte Carlo events  
and repeating the fit procedure.  Some analysis cut variations 
significantly increased the level of backgrounds to the 
$K_{L}\rightarrow\pi^{+}\pi^{-}e^{+}e^{-}$ mass peak.  
These cuts were separated from the other physics cuts and are labled as ``background'' in 
Table~\ref{systematics}.  Finally, input parameters to the Monte Carlo
such as $\eta_{+-}$,$\Phi_{+-}$, and the strong interaction
$\pi^+\pi^-$ phases shifts $\delta_{0,1}$ that were not included in the
liklihood fit were varied by $\pm 1 \sigma$ of
their published values to determine the uncertainty in the fit parameters due
to their uncertainties.   The total systematic errors in $a_1$/$a_2$, 
$|\tilde{g}_{M1}|$, $|g_{CP}|$ and $\frac{|g_{E1}|}{|g_{M1}|}$ were 
obtained by adding the systematic errors in quadrature.

The systematic errors in the $\phi$ asymmetry due to several sources
are given in Table~\ref{systematics2} below. The physics cut variations 
and background systematics of the $\phi$ angle asymmetry 
have been determined as discussed above. The $\eta_{+-}$, $\Phi_{+-}$ and 
$\delta_{0,1}$  systematics are obtained as before using the
$\pm 1\sigma$ uncertainties in these parameters.  Additional uncertainties of
the asymmetry due to the one $\sigma$ uncertainties of the fitted parameters
are also included.  All systematic errors of Table~\ref{systematics2} are added 
in quadrature to obtain the total systematic error.

In conclusion, the KTeV collaboration measured a charge radius coupling 
$|g_{CP}|=0.163\pm0.014({\rm stat})\pm0.023({\rm syst})$ which has been used to obtain, 
in a novel way~\cite{ref:4}, a $K^0$ charge radius of $\langle r^2_{K^0} \rangle =
-3|g_{CR}|/M^2_K = 
(-0.077\pm0.007({\rm stat})\pm0.011({\rm syst})) (fm^2)$, consistent with the previous measurements of the
$K^0$ charge radius~\cite{ref:15,ref:16,ref:17} obtained in $K^0$ electron scattering 
and from a similar analysis of the $K_{L}\rightarrow\pi^{+}\pi^{-}e^{+}e^{-}$ mode
by NA48~\cite{ref:18}.  We also set a
first experimental upper limit on the presence of E1 direct photon emission
in the $K_{L}\rightarrow\pi^{+}\pi^{-}e^{+}e^{-}$ mode of 
$\frac{|g_{E1}|}{|g_{M1}|}<0.04$ (90\%CL) including systematic errors.  
In addition, the M1 photon emission coupling was measured to be 
$|\tilde{g}_{M1}| = 1.11\pm{0.12}~({\rm stat})\pm 0.08~({\rm syst})$
plus a vector form factor as given in equation (2) with $a_1$/$a_2 = (-0.744\pm 0.027~({\rm stat})\pm
0.032~({\rm syst}))~{\rm GeV}^2/c^2$. Using $a_1$/$a_2$ and $|\tilde{g}_{M1}|$, an average
$|{g}_{M1}|$ over the range of $E_{\gamma^*}$ was calculated to be $0.74\pm0.04$.
Finally, we made a measurement of a large CP-violating asymmetry in the
distribution of T-odd angle $\phi$ in $K_{L}\rightarrow\pi^{+}\pi^{-}e^{+}e^{-}$
decays of $(13.6\pm 1.4~({\rm stat})\pm 1.5~({\rm syst}))$\%
consistent with the theoretically expected asymmetry of Refs.~\cite{ref:4,ref:5}.  
This result is consistent with our original measurement~\cite{ref:2} and a later
measurement by NA48~\cite{ref:18}.  

We thank the Fermilab staff and the staffs of the participating institutions for their vital contributions.
This work was supported by the U.S. Department of Energy, the U.S. National Science Foundation, 
the Ministry of Education and Science of Japan, the Fundao de Amparo a
Pesquisa do Estado de So Paulo-FAPESP, the Conselho Nacional de
Desenvolvimento Cientifico e Tecnologico-CNPq, and the CAPES-Ministerio da Educao. 

\vspace{0.25cm}
\noindent $^{\dagger}$ To whom correspondence should be addressed.\\
\noindent Electronic address: ag@fnal.gov \\
\noindent $^{*}$Permanent address C.P.P. Marseille/C.N.R.S., France\\

\begin{table}
\begin{tabular}{|l|c|c|c|c|}
                                   &
\multicolumn{4}{|c|}{Uncertainty in} \\
\cline{2-5}
Source                             & $a_1/a_2$ & $|\tilde{g}_{M1}|$  & $|g_{CR}|$ &  $\frac{|g_{E1}|}{|g_{M1}|}$    \\
\hline
Monte Carlo Statistics              & 0.002     &  0.01              &   0.001    &   0.001  \\
Choice of initial MC parameters     & 0.005     &  0.02              &   0.001    &   0.001  \\
Skewing from input MC values        & 0.000     &  0.028             &   0.002    &   0.010  \\ 
Physics cut variations              & 0.022     &  0.041             &   0.021    &   0.018  \\
Background                          & 0.022     &  0.05              &   0.01     &   0.008  \\
$\eta_{+-}$ Uncertainty             & 0.0001    &  0.01              &   0.002    &   0.0002  \\
$\Phi_{+-}$ Uncertainty             & 0.0003    &  0.002             &   0.0002   &   0.0005  \\
$\delta_{0,1}$ Uncertainty          & 0.001     &  0.004             &   0.001    &   0.0003  \\
\hline
Total Systematic Error              & 0.032     &  0.08              &   0.023    &   0.023   \\
\end{tabular}
   \caption{Syst. errors of $a_1$/$a_2$, $|\tilde{g}_{M1}|$, 
$|g_{CR}|$ and  $\frac{|g_{E1}|}{|g_{M1}|}$}
 \label{systematics}
\end{table}

\begin{table}
\begin{tabular}{|l|c|}
Source                      & $\Delta$ Asymmetry (\%) \\
\hline                                        
Physics cut variations      & 0.71     \\
Background                  & 0.30     \\
$\eta_{+-}$ Uncertainty     & 0.163    \\
$\Phi_{+-}$ Uncertainty     & 0.111    \\
$\delta_{0,1}$ Uncertainty  & 0.325    \\
$\frac{|g_{E1}|}{|g_{M1}|}$ & 0.326    \\
$|g_{M1}|$,$a_1/a_2$        & 0.335    \\
$|g_{CR}|$                  & 0.335    \\
\hline                                          
Total Systematic Error      & 1.46     \\
\end{tabular}
   \caption{ CP violating $\phi$ asymmetry systematic errors}
 \label{systematics2}
\end{table}


\begin{thebibliography}{99}
\bibitem{ref:4} P. Heiliger and L.M. Sehgal, Phys. Rev. \textbf{D48},
4146(1993); L.M. Sehgal and M. Wanninger, Phys. Rev. \textbf{D46},
1035(1992); {\em ibid.} \textbf{D46}, 5209(E)(1992). 
\bibitem{ref:1} J.Adams {\em et al.}, Phys. Rev. Letters \textbf{80},
4123(1998).
\bibitem{ref:2} A. Avati-Harati {\em et al.}, Phys. Rev. Lett. \textbf{84},
408(2000).
\bibitem{ref:13} A. Golossanov, PhD Thesis, Univ. of Virginia, (2005). 
\bibitem{ref:5} J.K. Elwood {\em et al.}, Phys. Rev. \textbf{D52},
5095(1995);  J.K. Elwood {\em et al., ibid.}, \textbf{D53},
2855(E)(1996); J.K. Elwood {\em et al., ibid.}, \textbf{D53},
4078(1996).
\bibitem{ref:6a} E. Barberio and Z. Was, Comput. Phys. Commun. \textbf{79}, 291(1994).
\bibitem{ref:6} E.J. Ramberg {\em et al.}, Phys. Rev. Lett. \textbf{70},
2525(1993).
%\bibitem{ref:7} K. Hagiwara {\em et al.}, Phys. Rev.\textbf{D66}, 1(2002).
\bibitem{ref:phases} S. Pislak {\em et al.}, Phys. Rev. Lett. \textbf{87}, 221801(2001);
G. Colangelo  {\em et al.},, Nucl. Phys. \textbf{B603}, 125(2001).
\bibitem{ref:15} H. Foeth {\em et al.}, Phys. Lett. \textbf{B30}, 276(1969).
\bibitem{ref:16} F. Dydak {\em et al.}, Nucl. Phys. \textbf{B102}, 253(1976).
\bibitem{ref:17} W.R. Molzon {\em et al.}, Phys. Rev. Lett. \textbf{41}, 1213(1970).
\bibitem{ref:18} A. Lai {\em et al.}, Eur. Phys. J., \textbf{C30}, 33(2003).
%\bibitem{ref:9} I.I. Bigi and A.I. Sanda, hep-ph/9904484, (April, 1999).
\end{thebibliography}
\end{document}